# Optical properties along the c axis of $YBa_2Cu_3O_{6+x}$, for x = 0.50 → 0.95: evolution of the pseudogap

C.C. Homes[*] and T. Timusk[**]

*Department of Physics and Astronomy, McMaster University Hamilton, Ontario, Canada L8S 4M1*

D.A. Bonn, R. Liang and W.N. Hardy

*Department of Physics, University of British Columbia Vancouver, B.C., Canada V6T 2A6*

The reflectance of high-quality, unpolished single crystals of $YBa_2Cu_3O_{6+x}$, for the doping range $x = 0.50 \to 0.95$, has been measured with radiation polarized along the c axis from $\approx 50$ cm$^{-1}$ to 5000 cm$^{-1}$ between 10 K and 300 K. In highly doped ($x = 0.95$) material, the normal-state conductivity shows a metallic response. For intermediate dopings ($x = 0.85 \to 0.90$) the conductivity is no longer metallic, increasing with temperature, and for low dopings ($x = 0.50 \to 0.80$) this behavior is clearly seen to be caused by a pseudogap at $\approx 280$ cm$^{-1}$, which develops well above $T_c$. In the superconducting state, in the optimally-doped material, a gap-like depression develops below $T_c$ but there is residual conductivity to very low frequency indicating either an anisotropic gap, or a gap with nodes. In the underdoped materials the superconductivity-induced changes to the conductivity are harder to see, but it appears that the weight of the condensate originates at frequencies much higher than the conventional weak-coupling value of $3.5k_BT_c$.

## I. INTRODUCTION

Our current understanding of the optical conductivity in the high-temperature superconductors is limited. [1,2] The conductivity is anisotropic, with Drude-like free carrier transport in the copper-oxygen planes, while normal to the planes the transport varies from metallic at optimal-doping levels to nearly insulating at reduced-doping levels where the resistivity increases with decreasing temperature. The c-axis conductivity of $YBa_2Cu_3O_{6+x}$, the most studied material, [3–10] shows a metallic background with a very large scattering rate [3] and it is generally agreed that the transport is incoherent — that is the mean-free path appears to be less than a lattice spacing. The magnitude of the reported background conductivity varies with doping, ranging all the way from essentially insulating, [11] at low doping levels, to a low of 25 $\Omega^{-1}$cm$^{-1}$ for oxygen-reduced materials with $T_c = 60$ K. [12] to a conductivity as high of 450 $\Omega^{-1}$cm$^{-1}$ for overdoped crystals [10] at 100 K. The higher conductivity levels are correlated with higher doping levels. The optimally-doped material seems to have a dc conductivity in the neighborhood of 200 $\Omega^{-1}$cm$^{-1}$.

Superimposed on this electronic background are five strong phonon lines. These five lines are also seen in the spectra of ceramic superconductors, [13,14] and have been studied by a large number of groups. The lines which are quite sharp in the highly-doped material, display several unusual properties as the oxygen content is reduced. The phonon contribution to the conductivity will be discussed more fully in a separate publication. [15]

In the fully-doped materials, in the superconducting state well below $T_c$, a gap-like depression developes in the conductivity with a characteristic energy scale of $\approx 230$ cm$^{-1}$, or $2\Delta = 3.5k_BT_c$ (where $\Delta$ is the familiar BCS energy gap). In contrast to the fully-doped case, no clear gap-like feature which can be associated with superconductivity is seen below $T_c$ in the underdoped materials, [9] and the spectral weight for the condensate seems to come from very high frequencies, corresponding to an energy scale of the order of 1000 cm$^{-1}$. However, in the underdoped materials there is a well-defined depression in the conductivity below $\approx 200$ cm$^{-1}$ which is observed well above $T_c$; this normal-state gap-like feature is referred to as a *pseudogap*.

A clear pseudogap has also been reported the double-chain $YBa_2Cu_4O_8$, [16] and there is also evidence of one in $Pb_2Sr_2RCu_3O_8$. [17] A temperature-dependent depression in the low-frequency conductivity in the normal state can also be seen in $La_{2-x}Sr_xCuO_4$, [18] which has been interpreted recently as evidence of a pseudogap. [19,20]

In this paper we report on the c-axis polarized reflectance of a whole range of dopings of $YBa_2Cu_3O_{6+x}$, ranging from $x = 0.50$ to 0.95 of large, high-quality single crystals as a function of temperature, both in the normal and in the superconducting state. The improved sample quality and size has allowed us to examine in some detail the c-axis properties of this anisotropic system by systematically following the effects of oxygen doping. Pre-liminary reports of this work have been published. [9,21]

## II. EXPERIMENTAL

### A. Sample Preparation

Large, $YBa_2Cu_3O_{6+x}$ single crystals, grown by a flux method in yttria-stabilized zirconia (YSZ) crucibles, [22] were used in the optical measurements. High-purity $Y_2O_3$, CuO and $BaCO_3$ were used as starting materials and were mixed in a power mortar with agate components. The mixed powder was pressed into a pellet and placed into a YSZ crucible. The crucible was first heated to 850°C and held there for 4 h to evaporate carbonates. The temperature was then increased to 1010°C, held for 4 h and cooled to 990°C. After soaking at 990°C for $8-16$ h, the crucible was slowly cooled at a rate of $0.3-0.8$°C/h to $950-970$°C. The flux was then poured onto a porous ceramic. Finally, the crucible was cooled to room temperature at a rate of 150°C/h. Assays by ICP mass spectroscopy indicate impurities at the 0.1% level.

The as-grown crystals had an orthorhombic lattice symmetry and show a broad superconducting transition with an onset at 85 K. They were subsequently annealed in a furnace for seven days in oxygen and the oxygen content was initially set to 6.95 by the use of the empirical relationship between annealing temperature, oxygen partial pressure and $x$. [23] This oxygen content yields a $T_c$ of 93.2 K and outstanding bulk homogeneity which is demonstrated by a specific heat jump at $T_c$ that is very narrow, only 0.2 K wide. The transition width, as measured by resistivity was 0.25 K (taken from the onset to zero resistivity). The crystals were deoxygenated by annealing at 650 °C under vacuum. Once the amount of deoxygenation has been determined using volumetric techniques, [23] the sample is annealed at 650°C for several days, and then slowly cooled to room temperature. A total of three crystals were examined; the first had an initial value of $x = 0.95$, and was subsequently deoxygenated to 0.90 and 0.85. The other two crystals had oxygenations of $x = 0.95$ and 0.70, and 0.60 and 0.50, respectively. The oxygen contents and the $T_c$'s have been summarized in Table I. The typical dimensions of the face containing the c axis was $\approx 1.5 \times 0.5$ mm.

### B. Reflectance measurements

The sample was glued to the apex of a cone, oriented so that light is incident on a face of the crystal that contains the c-axis. To utilize the whole sample, an overfilling technique is used so that light that misses the sample is scattered out of the optical path. [24] The reflectance of the sample ($R_s$) is compared to the reflectance of a stainless-steel reference mirror ($R_m$). To correct for the sample size and any irregularities in the surface, and to eliminate the effects of the reference mirror, the sample was overcoated with gold *in situ* and the measurements were repeated on the gold-coated sample ($R_{gs}$). The effects of the reference mirror may be removed by dividing the two ratios

$$\left(\frac{R_s}{R_m}\right)\left(\frac{R_{gs}}{R_m}\right)^{-1} = \frac{R_s}{R_{gs}},$$

which yields the reflectance of the sample with respect to gold. The optical properties of gold are well known, [25] and the reflectance may subsequently be corrected by multiplying the ratio by the reflectance of gold to yield the absolute reflectance of the sample. The estimated accuracy of the absolute reflectance is estimated to be $\pm 1\%$.

TABLE I. The critical temperatures of $YBa_2Cu_3O_{6+x}$ for $x = 0.50 \to 0.95$. The low-temperature values (10 K) of the spectral weight of the condensate ($\omega_{ps}$), determined from an analysis of $\epsilon_1$, (discussed in the text), and the London penetration depth ($\lambda_L$), are also listed.

| $x$ | $T_c$ (K) | $\omega_{ps}$ (cm$^{-1}$) | $\lambda_L$ ($\mu$m) |
|---|---|---|---|
| 0.95 | 93.2 | 1497 | 1.06 |
| 0.90 | 91.5 | 1003 | 1.58 |
| 0.85 | 89   | 786  | 2.03 |
| 0.80 | 78   | 472  | 3.37 |
| 0.70 | 63   | 308  | 5.17 |
| 0.60 | 58   | 244  | 6.52 |
| 0.50 | 53   | 204  | 7.80 |

The brass cone holding the sample was anchored with copper braid to the cold finger of a flow dewar. Samples mounted in cold-finger dewars that are illuminated with ambient radiation are subject to heating. We tested the sample temperature by mounting a thermometer in place of the sample. This test suggested that the temperature of the sample is within 2 K of the temperature of the cold finger. We also used the superconducting transition of the sample itself as an internal test and found that, for the $x = 0.95$ sample, the optical properties began to change rapidly below a thermometer reading of $93 \pm 2$ K, in excellent agreement with the transition temperature measured by magnetic susceptibility on the sample of 93.2 K. At very low temperatures (4.2 K), the sample tends to be slightly warmer the tip of the cold finger; we estimate our lowest measured temperature at the sample to be $10 \pm 4$ K.

The orientation of the c axis was determined by plotting the reflectance ratio as a function of the polarizer angle, and then least-squares fitting to a $\sin^2 \theta$ function. The estimated error in the angle between the c axis of the crystal and the polarizer axis is less than 1°. The polarizer leakage of other polarizations is less than 0.4% of the signal.

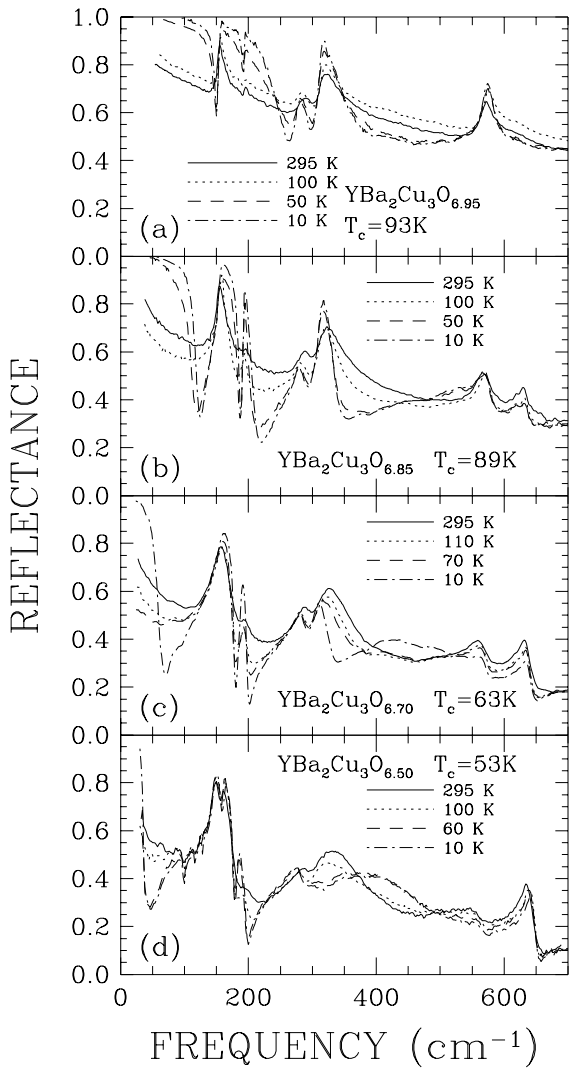

FIG. 1. The reflectance of $YBa_2Cu_3O_{6+x}$ for radiation polarized along the c axis from $\approx 50$ cm$^{-1}$ to 700 cm$^{-1}$ at several temperatures above and below $T_c$, for four oxygen dopings $x$,(a) 0.95, (b) 0.85, (c) 0.70 and (d) 0.50. The increase in the reflectance of the highly-doped material indicates a metallic response, whereas the drop in the reflectance in the materials with lower oxygen dopings indicates a non-metallic type of behavior. At low frequencies, the normal-state reflectance always approaches unity, indicating a non-zero conductivity instead of insulating behavior. Below $T_c$ the curvature of the reflectance changes, indicating the formation of a zero-frequency condensate.

## III. RESULTS

### A. The c-axis reflectance

The reflectance for light polarized parallel to the c axis is shown in Fig. 1 from $\approx 50$ cm$^{-1}$ to 700 cm$^{-1}$ for $YBa_2Cu_3O_{6+x}$, for $x = 0.50, 0.70, 0.85$ and $0.95$ at several temperatures above and below $T_c$. The overall reflectance is typical of marginally-metallic systems: starting near unity at low frequency, the average reflectance drops with frequency except at the sharp resonances due to phonons. As the temperature is reduced the reflectance of the fully-doped material ($x = 0.95$) increases, characteristic of normal metallic temperature behavior. In the oxygen-reduced curves the opposite is true, the reflectance decreases as the temperature is lowered. This is in accord with the non-metallic behavior of the c-axis dc conductivity in oxygen-reduced samples.

Below the superconducting transition, dramatic changes take place in the reflectance of all the samples: a region of nearly unit reflectance appears at low frequency immediately below $T_c$. A characteristic plasma-like edge appears at $\approx 250$ cm$^{-1}$ in the fully-doped sample and its frequency diminishes with doping. This plasma edge is associated with the presence of a zero-frequency condensate. [26] The shape of the two low-frequency phonons is also altered, a sign that the background continuum conductivity has reduced markedly. However, it is clear from the reflectance that no true superconducting gap has formed in the region of the plasma edge, since the reflectance does not go to zero as expected for a plasma edge in the absence of damping. This follows from the relationship between the reflectance and the conductivity. If we define $\omega_1$ as the frequency, close to the minimum of the reflectance at the plasma edge, where the index of refraction $n = 1$, we have at $\omega_1$:

$$\sigma_1(\omega_1) = \frac{\omega_1}{\pi}\left[\frac{R_1}{1-R_1}\right]^{1/2}, \qquad (1)$$

where $R_1$ is the reflectance at $\omega_1$, then if $R_1 \neq 0$, $\sigma_1(\omega_1) \neq 0$ as well.

The optical constants have been calculated by a Kramers-Kronig transformation of the reflectance, and the optical conductivity corresponding to the reflectance curves of Fig. 1 are shown in Fig. 2. The transformations were performed by extending the reflectance as follows: in the normal state the reflectance below the lowest measured frequency was assumed to follow Hagen-Rubens frequency dependence: $(1 - R) \propto \omega^{-1/2}$. In the superconducting state, where a region of near unit reflectance develops at low frequency, the extrapolation was done assuming that $(1 - R) \propto \omega^2$. This form is equivalent to assuming that there is a finite constant conductivity component down to zero frequency. The alternate assumption that there is a gap, starting at a frequency where our signal-to-noise ratio cannot distinguish between unit reflectance and a small amount of absorption, gives only slightly different results in the region of actual valid data. Above 5000 cm$^{-1}$ the measurements of Koch et al. [5] were used up to $3.5 \times 10^5$ cm$^{-1}$; above this frequency free-electron behavior was assumed ($R \propto \omega^{-4}$).

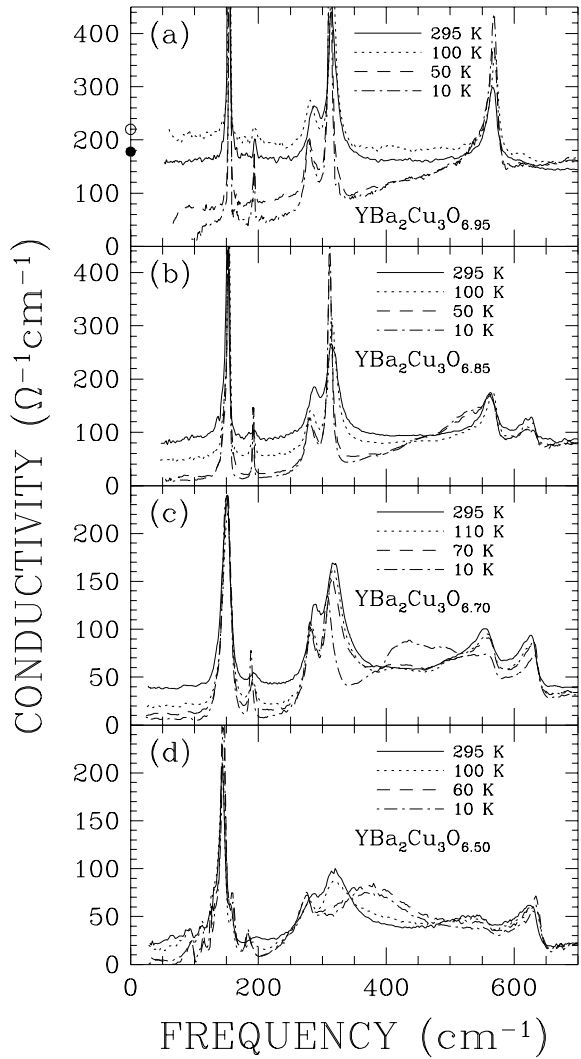

FIG. 2. The optical conductivity ($\sigma_1$) for $YBa_2Cu_3O_{6+x}$ for radiation polarized along the c axis from $\approx$ 50 cm$^{-1}$ to 700 cm$^{-1}$ at several temperatures above and below $T_c$, for four oxygen dopings $x$ (a) 0.95, (b) 0.85, (c) 0.70 and (d) 0.50. In the highly-doped material, below $T_c$ (93.5 K), a gap-like depression opens up below $\approx$ 500 cm$^{-1}$, but at 10 K the conductivity is non-zero down to at least 100 cm$^{-1}$. In the $x$ = 0.85 material, the conductivity is clearly non-metallic in nature, a trend which resolves itself into a pseudogap for $x \leq$ 0.70. The solid and open circles shown in the top panel are the c-axis conductivity as determined by four-probe measurements (Ref. 29) at room temperature, and just above $T_c$.

### B. The phonon spectrum

In the highly-doped material, closest to the true orthorhombic phase of $YBa_2Cu_3O_7$, there are five strong phonons; at low temperature their frequencies are: 155, 194, 277, 312 and 570 cm$^{-1}$. The 570 cm$^{-1}$ line has a Lorentzian shape at room temperature, but at low temperature it becomes asymmetric [27] and is usually described in terms of a Fano line shape: $\sigma_1(\omega) = A[(x+q)^2/(1+x^2)]$, where $\sigma_1(\omega)$ is the conductivity, $A$ is a constant, $x = (\omega - \omega_i)/\gamma_i$, $\omega_i$ is the phonon frequency, $\gamma_i$ is the phonon line width and $q_i$ is a parameter that describes the asymmetry of the $i$th phonon. [28] While the Fano line shape describes the 570 cm$^{-1}$ at low temperatures, it fails as the line shape becomes more Lorentzian ($q \to \infty$).

A better fit to the data is in terms of a classical Drude-Lorentz model for the dielectric function, but with an empirical line shape made up of a mixture of the real and imaginary parts of Lorentzian oscillators:

$$\tilde{\epsilon}(\omega) = \epsilon_\infty + \sum_j \frac{\omega_{pj}^2 e^{i\theta_j}}{\omega_{TO,j}^2 - \omega^2 - i\omega\gamma_j}, \quad (2)$$

where $\epsilon_\infty$ is the contribution to the real part of the dielectric function at high frequency, and $\omega_{pj}$ is the effective plasma frequency, $\omega_{TO,j}$ and $\gamma_j$ are the center frequency and the damping constant, and $\theta_j$ is the phase associated with the $j$th phonon. The line shape is that of a classical oscillator for $\theta = 0$, but becomes asymmetric for nonzero $\theta$. An advantage of this line shape is that, unlike in the case of the Fano shape, an oscillator strength can be associated with it. The degree of mixing is a measure of the asymmetry and may be related approximately to the Fano parameter $q$ by $q^{-1} \propto \tan(\theta/2)$. All the phonon lines have been fitted to Eq. 2 with a non-linear least-squares method. In addition to the parameters of the modified-Lorentzian oscillator, a linear background was used in the fits. We find that within the accuracy of our experiments, the four low-frequency lines are symmetric, and only the 570 cm$^{-1}$ line has clearly asymmetric shape, which becomes more pronounced at low temperatures.

An artificial asymmetry is introduced if an incorrect value for the 100% line is used in the spectra. The lines that have the highest reflectance are most susceptible to this error, which in our case is the 155 cm$^{-1}$ line, which we find accurately symmetric, thus confirming the position of our 100% line. Another test of the accuracy of the conductivity is to compare the optically determined conductivity to the dc measurements on samples from the same batch. Tests done on the fully-doped sample show that the optical conductivity agrees with the dc conductivity to within 5%.

### C. The c-axis optical conductivity

The two components of the conductivity along the c-axis can be separated: there is a continuous background electronic conductivity that increases rapidly in magnitude as the doping level increases towards full doping, and a series of sharp phonon bands. In the underdoped materials two important changes to the phonons take place: (i) a new phonon appears $\approx$ 610 cm$^{-1}$, and (ii) the spectral weight of the several high-frequency phonons is transferred into a new broad band centered at ≈ 400 cm$^{-1}$ — this process begins in the normal state, but is most prominent at low temperature.

The continuous background conductivity in the normal-state of the $x = 0.95$ material is nearly frequency independent at room temperature, but shows a slightly metallic character in that the conductivity increases as the temperature is lowered and there is small Drude-like decrease in the conductivity with frequency as well. In agreement with this, dc resistivity of the fully-doped material has a positive slope with temperature. The extrapolated values of the dc conductivity, for this doping level, agree (to within 5%) with direct dc measurements of the on crystals from the same batch, [29] shown in Fig. 2 by the open and closed circles.

If the doping is decreased below $x = 0.90$, major changes take place in the normal-state conductivity. The most striking is the development of a pseudogap below the region of ≈ 200 cm$^{-1}$ which can be seen even in the presence of the phonons, as shown in Fig. 2. To illustrate the development of the pseudogap with doping we plot in Fig. 3(a) the conductivity at 50 cm$^{-1}$, below the gap frequency at 295 and 10 K, respectively. The inset in Fig. 3(a) shows the corresponding room-temperature resistivity at ≈ 50 cm$^{-1}$ as a function of doping in the $x = 0.50 \to 0.95$ region (the values below $x = 0.5$ are adapted from Fig. 5 in Ref. 2); the resistivity is increasing exponentially with decreasing doping. The crossover to metallic behavior where the conductivity decreases with temperature occurs at $x \approx 0.92$. Below this concentration the low-temperature conductivity (closed circles) is depressed as the result of the development of the pseudogap, and it is clear from Fig. 2 that the temperature dependence of the low-frequency conductivity in the c-direction in oxygen-reduced YBa$_2$Cu$_3$O$_{6+x}$ is directly related to the depth of the pseudogap. This is illustrated in Fig. 4 where we plot, as function of temperature, the extrapolated low-frequency conductivity compared with four-probe measurements of Baar *et al.* [29] on crystals from the same batch and with measurements of Iye *et al.* [12] for three other doping levels.

The agreement between the infrared and dc measurements is excellent in both absolute value and temperature dependence for materials at the extreme ends of the doping range, ($x = 0.60$ and $0.95$, respectively). The agreement at intermediate dopings is not as good. For example the 80 K curve of Iye *et al.* agrees better with our 89 K curve than with the 78 K one. While it is possible that there is a genuine discrepancy between the dc and infrared conductivities at intermediate doping levels, it should be noted that properties change rapidly with doping in this range and there could be some uncertainty in the exact oxygen content as well as the homogeneity of the doping in region between the fully-doped materials where all the chains are occupied and the alternating-chain compounds in the T$_c$ = 60 K plateau region.

With these reservations at intermediate doping levels, we conclude, from the good agreement between the mag-nitude of the dc conductivity and the far-infrared conductivity in the pseudogap, that the temperature dependence of the c-axis dc conductivity in the normal state is governed by the depth of the pseudogap. Well above the frequency of the pseudogap at 800 cm$^{-1}$, shown in Fig. 3(b) the c-axis conductivity is *temperature independent* at all doping levels, except the most metallic.

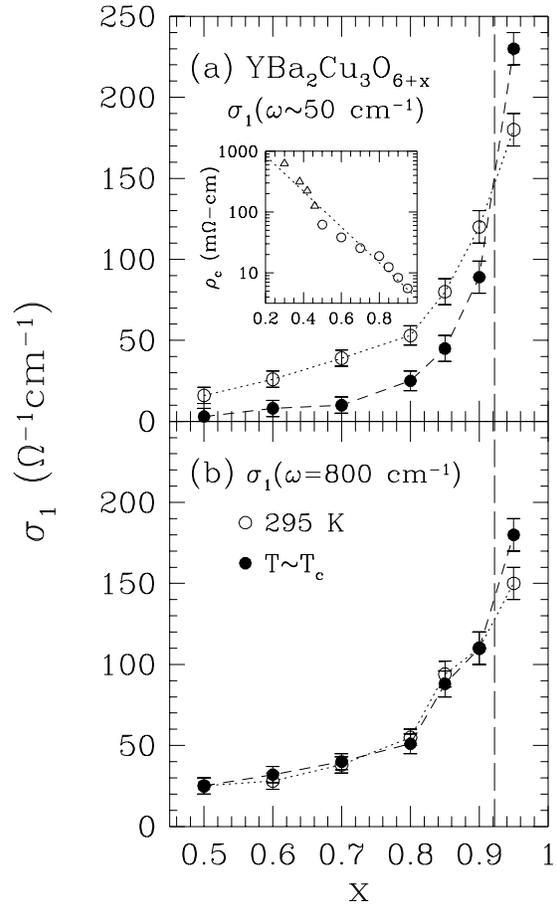

FIG. 3. The normal-state optical conductivity of YBa$_2$Cu$_3$O$_{6+x}$ as a function of oxygen doping at (a) 50 cm$^{-1}$ at 295 K (open circles), and just above $T_c$ (solid circles), and (b) at 800 cm$^{-1}$ at 295 K (open circles) and just above $T_c$ (solid circles). While the high-frequency conductivity shows little temperature dependence, the low-frequency conductivity is very sensitive to both doping and temperature and shows a crossover from "metallic" to "non-metallic" behavior at $x \approx 0.92$. The value of the conductivity at this point (≈ 150 $\Omega^{-1}$cm$^{-1}$) is near the Mott limit for metallic conductivity in these materials. The inset in (a) shows the exponentially increasing at resistivity as a function of doping at room temperature; the circles are from this work, and the triangles showing dopings below $x = 0.50$ are adapted from Fig. 5 in Ref. 2. The dotted line in the inset is the linear fit to the data.

In the superconducting state the fully-doped material shows rapid changes in the conductivity just below $T_c$. In contrast to the normal state where the conductivity is *increasing* as the temperature is lowered, below $T_c$ the low-frequency conductivity begins to *decrease* and a gap-like depression develops with a characteristic energy of $\approx 200$ cm$^{-1}$. Qualitatively this is reminiscent of BCS dirty-limit superconductivity: metallic in the normal state but showing a decreasing conductivity below the superconducting transition as a gap develops. The conductivity drops to less than 5% of the normal-state value at the lowest temperature and the lowest frequency, but no true gap can be seen down to $\approx 100$ cm$^{-1}$ (the lower limit of our sensitivity).

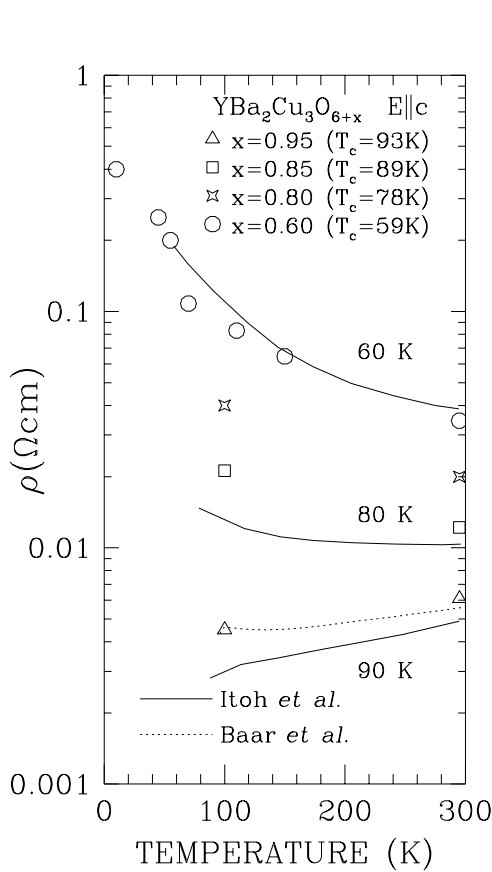

FIG. 4. The extrapolated value of the resistivity along the c-axis of $YBa_2Cu_3O_{6+x}$ at $\approx 50$ cm$^{-1}$ for the four doping levels $x = 0.60, 0.80, 0.85$ and $0.95$ (open shapes), compared to the four-probe c-axis resistivity measurements on single crystals by Iye *et al.* [Ref. 12] (solid line) and Baar *et al.* [Ref. 28] (dotted line). The extrapolated values for the $x = 0.60$ and $0.95$ dopings agree quite well with the four-probe measurements. However, the agreement is poorer for intermediate dopings, where the physical properties of the materials are changing rapidly with doping.

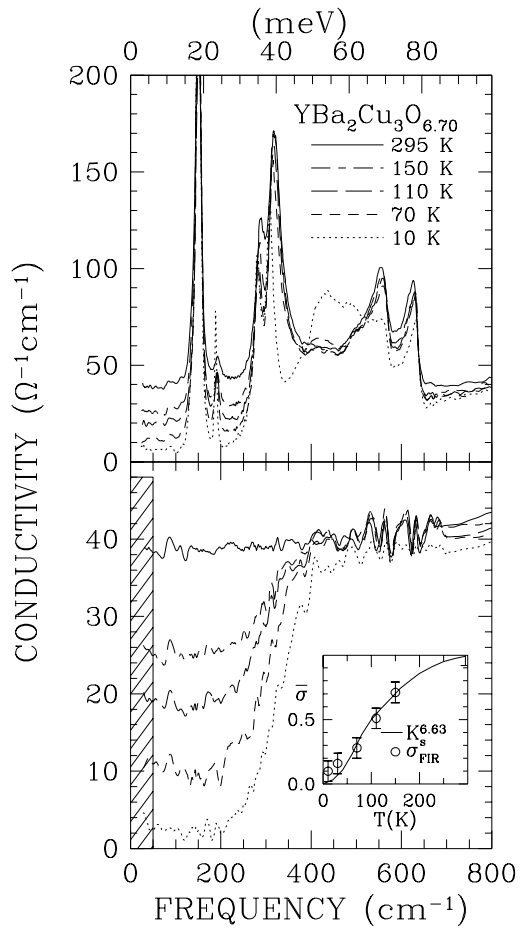

FIG. 5. The optical conductivity along the c-axis $YBa_2Cu_3O_{6.70}$ ($T_c$=63 K) at 295, 150, 110, 70 and 10 K with (a) phonons present, and (b) the phonons removed. As well as the five strong $B_{1u}$ phonons, the feature at $\approx 400$ cm$^{-1}$ (believed to be due to phonons) has also been removed. The conductivity in (b) shows the conductivity somewhat suppressed at room temperature — a clearly defined gap continues to develop in the normal state, a trend which continues into the superconducting state.

At low-doping levels the behavior in the superconducting state is very different. While the presence of superconducting condensate is seen clearly in the reflectance, through the formation of a plasma edge, changes to the real part of the conductivity are less dramatic than those in the fully-doped material. The pseudogap dominates the behavior of the conductivity at low frequency and a plot of the low-frequency conductivity as a function of temperature drops continuously through $T_c$, with no obvious break at the superconducting transition, such as what is seen in the fully-doped material. [9]

Nevertheless, a careful examination of the conductivity reveals subtle changes on entry into the superconducting state. The first of these is the transfer of spectral weight from finite frequencies to the condensate delta function at zero frequency. This can be seen better in the spectra of the naturally underdoped $YBa_2Cu_4O_8$ material where the condensate spectral weight is very large, [16] and there is reduction of conductivity in the superconducting state over a broad frequency range, $\approx 1000$ cm$^{-1}$. To see more clearly the changes in the conductivity at $T_c$ it is necessary to first suppress the phonons.

We use the phonon parameters determined by the least-squares fits to suppress the phonon structure by subtracting the calculated phonon conductivity from the overall conductivity spectra. Where the phonon lines are sharp and well isolated, such as in the pseudogap range for frequencies $\lesssim 200$ cm$^{-1}$, this procedure can be done quite well. However, in the $\approx 400$ cm$^{-1}$ region of the underdoped materials, the appearance of a new feature (or set of features) in this region in the normal state, and increasing strength at low temperatures, makes the procedure for phonon subtraction difficult. This situation creates some uncertainty about the nature of the electronic background, as may be observed in Figs. 2(c) and 2(d). One possibility is that the broad feature at $\approx 400$ cm$^{-1}$ is electronic in origin and should therefore not be subtracted. However, the conductivity sum rule yields an important clue: we observe that the spectral weight of the phonons in the $200 - 600$ cm$^{-1}$ range, at room temperature where the phonons are sharp and well defined, is equal to the spectral weight at low temperature only if the broad peak at $\approx 400$ cm$^{-1}$ is included as a phonon. [15] Figures 5(a) and 5(b) show the conductivity for $YBa_2Cu_3O_{6.70}$ before and after the phonons have been removed in this way, respectively. Both of the panels in Fig. 5 show that between 70 K in the normal state, and 10 K in the superconducting state, a small uniform depression of conductivity of the order of 5 $\Omega^{-1}$cm$^{-1}$ over a very broad range of frequencies, similar to what is observed in the $YBa_2Cu_4O_8$ material. This depression seems to associated with the onset of superconductivity and not the pseudogap. A similar depression is seen in the $x = 0.60$ material [see Fig. 2(d)].

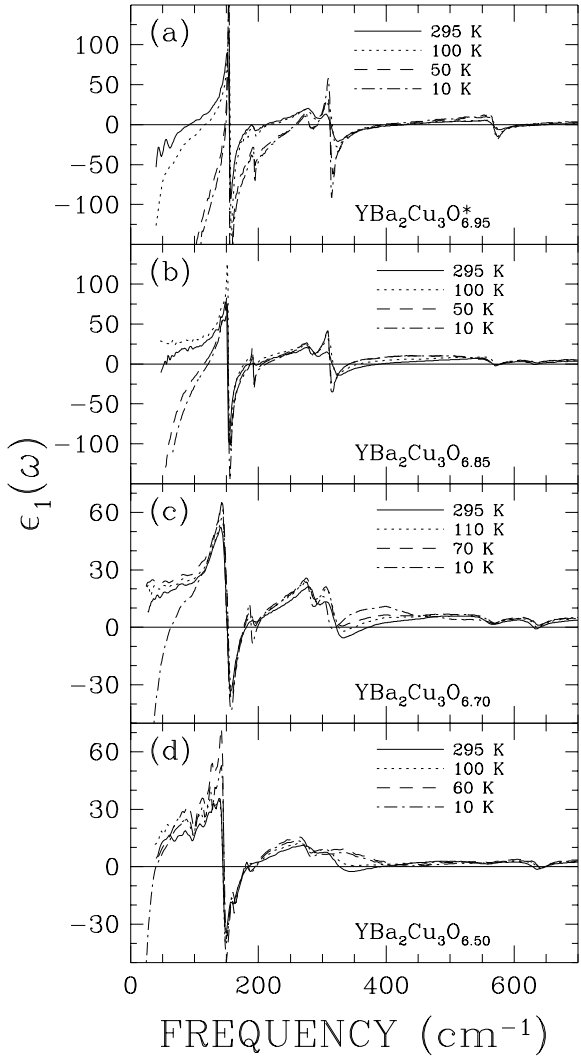

FIG. 6. The real part of the dielectric function ($\epsilon_1$) for $YBa_2Cu_3O_{6+x}$ for radiation polarized along the c-axis from $\approx 50$ cm$^{-1}$ to 700 cm$^{-1}$ at several temperatures above and below $T_c$ for the oxygen dopings $x = 0.60, 0.70, 0.85, 0.93$, and $0.95$. In the top panel, the two temperatures above $T_c$ are for $x = 0.93$, while the two below are for $x = 0.95$.

The real part of the dielectric function $\epsilon_1$, as calculated from the Kramers-Kronig analysis of the reflectance curves in Fig. 1, is shown in Fig. 6. The dispersion at low frequency, is dominated by the phonon at 150 cm$^{-1}$ which gives a positive contribution at low frequencies of the order of 20, and the large negative contribution of the condensate in the superconducting state. The phonon contribution is large enough to keep the net value of the low-frequency dielectric function positive in the normal state except for the most highly-doped material where a metallic negative response is seen at all temperatures.

At all other doping levels the low frequency $\epsilon_1$ is positive and constant in the normal state, the signature of non-metallic behavior, but it rapidly changes sign in the superconducting state where the condensate dominates the dispersion. The superconducting transition is clearly seen to lead to large and negative $\epsilon_1$ as $\omega \to 0$ for all doping levels. The interplay of the positive contribution of the phonons with the negative contribution from the condensate gives rise to a zero crossing of $\epsilon_1$ and a plasma edge in the conductivity spectra. The frequencies of the plasma edges correspond quite accurately to the zero crossing of $\epsilon_1$.

## IV. DISCUSSION

### A. The electronic conductivity

The conductivity along the c-axis direction in $YBa_3Cu_3O_{6+x}$ is very unusual. In the normal state, in the optimally and over-doped materials, the conductivity consists of a broad, Drude-like band with a high scattering rate.

As the doping level is lowered, the behavior of the normal-state conductivity changes dramatically: the overall conductivity becomes frequency and temperature independent and as the temperature is lowered and pseudogap develops with an onset of $\approx 200$ cm$^{-1}$.

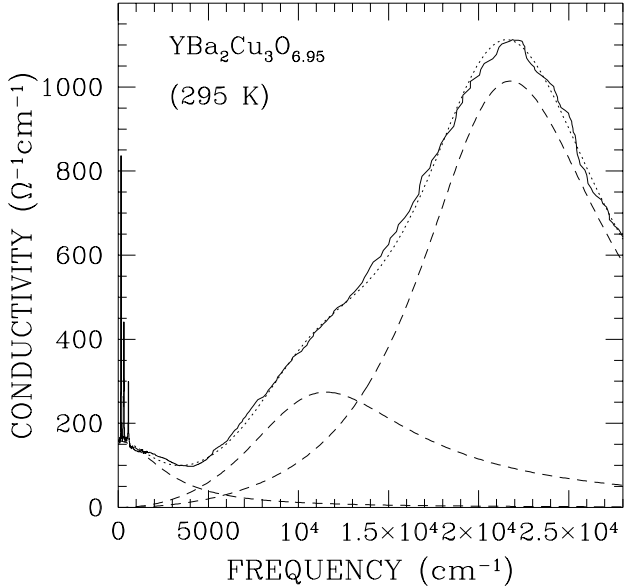

FIG. 7. The results of the Drude fit (dotted line) to the c-axis conductivity of $YBa_2Cu_3O_{6.95}$ at 295 K (solid line). The dashed lines show the components used in the fit: the Drude component centered at zero frequency, and the two oscillators in the near infrared and visible at $\approx 1.2$ and $2.2 \times 10^4$ $cm^{-1}$.

Since the conductivities in the fully-doped material ($x = 0.95$) and the under-doped samples ($x = 0.50 \rightarrow 0.80$) are so different, we will discuss these separately. The materials with intermediate doping ($x = 0.85 \rightarrow 0.90$) exhibited behavior that was intermediate between the fully doped and the under-doped samples and were not studied in great detail. It is possible that these samples were not homogeneous on the smallest scale, although they showed sharp transitions in magnetization, with no traces of multiple phases.

*1. Fully doped YBCO*

It is apparent that in the highly-doped materials the c-axis conductivity is becoming more metallic and may approach coherent transport in the overdoped case. [10] The simplest model that describes such transport is a Drude-Lorentz model for the dielectric function. In fitting to the Drude model, it is also important to take into consideration the effects of two high-frequency oscillators at 1.2 and $2.2 \times 10^4$ $cm^{-1}$ (in the near infrared and visible), as the long Lorentzian tails of these oscillators carry considerable weight in to the mid infrared. These two features have been observed in other c-axis studies, [2,5] and are relatively insensitive to the removal of oxygen. The fit using this approach is shown in Fig. 7 at 295 K, with the Drude parameters of $\omega_{pD} = 5230$ $cm^{-1}$ (0.66 eV), and $\Gamma_D = 2870$ $cm^{-1}$ (0.36 eV). The overall quality of the fit is quite good. Just above $T_c$ at 100 K, $\omega_{pD} = 5090$ $cm^{-1}$ (0.63 eV) and $\Gamma_D = 2010$ (0.24 eV). This yields scattering times ($\tau = \hbar/\Gamma_D$) of 11.6 fs and 16.6 fs at 295 K and 100 K, respectively. Using a band-structure Fermi velocity [30] of $v_F = 7 \times 10^6$ cm s$^{-1}$, yields mean-free paths of $\approx 9$ and 12 Å at 295 and 100 K. Both of these values are larger than the largest interatomic spacing along the c axis ($\approx 3.5$ Å), suggesting onset of coherent transport. This is consistent with the observation of a metallic response for this doping.

For $x = 0.90$, the mean-free path decreases to $\approx 4$ Å at $T_c$, and for lower oxygen dopings, it becomes less than an atomic spacing. Thus, the unphysical mean-free path and non-metallic behavior for $\lesssim 0.9$ rules out conventional Bloch-Grüneisen transport and implies an incoherent process. One model for incoherent transport is hopping conductivity between localized states, which is often seen in marginally-metallic systems. This mechanism leads to an approximately linear variation of conductivity with both frequency and temperature, [31–35] and is inconsistent with the negative Drude-like slope that we observe in the fully-doped material and the constant conductivity (at high temperature where the pseudogap has not yet formed) in the under-doped case.

Other approaches that have successfully described the ab-plane conductivity of the metallic high-temperature superconductors may also be applicable to the c-axis conductivity. The one-component model is expressed as a generalized Drude model, in which the damping rate is frequency dependent,

$$\tilde{\epsilon}(\omega) = \epsilon_\infty - \frac{\omega_p^2}{\omega[m^*(\omega)/m_b][\omega + i\Gamma^*(\omega)]} \quad (3)$$

where $\Gamma^*(\omega) = \hbar\omega\epsilon_2/(\epsilon_\infty - \epsilon_1)$ is the renormalized damping, and $m^*(\omega)/m_b$ is the effective mass enhancement over the band mass. Interpreted in this manner the low-frequency part of the conductivity would be attributed to a region where the scattering rate is low. By any estimate, the c-axis plasma frequency is considerably less than that for the ab-plane. As a result the zero-crossing of $\epsilon_1$ is quite low, and as it quickly approaches $\epsilon_\infty$ the renormalized damping becomes quite large. Thus, if $\epsilon_\infty \approx 4.5$, $\Gamma^* \approx 1.2 \times 10^4$ $cm^{-1}$ at 1000 $cm^{-1}$. The fact that the damping exceeds the frequency by a large factor is a sign that the transport is incoherent or diffusive.

In the two-component model of the conductivity, there are two channels of conductivity: the first is a coherent Drude component with a temperature-dependent damping, and the second consists of a broad mid-infrared component that is essentially temperature independent. Below $T_c$, the Drude component condenses, and all of its spectral weight is transferred to the delta function at the origin (the electromagnetic response of the superconducting condensate). Within this framework, the residual absorption is interpreted as being due to the mid-infrared component that does not participate in the formation of the superconducting condensate. The Drude contribution may be determined from the difference between the conductivity just above $T_c$, and a very low temperature. This amounts to assuming that the materials are in the clean limit.

In the initial fit using a simple Drude model, we included two high-frequency oscillators, but none in the mid-infrared. The temperature-dependent conductivity of $YBa_2Cu_3O_{6.95}$ shown in Fig. 2(a), implies that for the two-component model to be applied, at least one mid-infrared oscillator at $\approx 800$ cm$^{-1}$ must be included; two other bands at 2000 cm$^{-1}$ and 6000 cm$^{-1}$ have also been added (although their positions are somewhat arbitrary). The Drude-Lorentz model has been fit to the conductivity using a least-squares technique. Just above $T_c$, the Drude parameters are $\omega_{pD} = 2470$ and $\Gamma_D = 485$ cm$^{-1}$.

Below $T_c$, $\omega_{pD}$ is observed to decrease — this corresponds to the condensation of free carriers into the superfluid below $T_c$. As the Drude carriers are removed by this process, the background due to the mid-infrared bands is gradually revealed. The strength of the condensate at 10 K has been calculated from $\epsilon_1(\omega)$ to be $\approx 1500$ cm$^{-1}$, so that only a third of the free carriers condense in this model. Interestingly, while $\Gamma_D$ would be expected to decrease, it actually remains constant below $T_c$.

There are certain problems with the two-component picture. The first is in the calculation of the scattering rate from the dc conductivity. From the linear temperature dependence of the resistivity and a plasma frequency of $\approx 2500$ cm$^{-1}$ we write for the resistivity $\rho = 4\pi/\omega_{pD}^2\tau$, where

$$1/\tau = 1/\tau_0 + 2\pi\lambda k_B T,$$

and $1/\tau_0$ is the residual scattering rate, and $\lambda$ is the electron-boson coupling constant that is responsible for the temperature-dependent transport scattering. Applying this calculation to the c-axis dc transport data we find a $\lambda$ of the order of 0.1, similar to what is obtained in the ab plane. In this picture, it seems that the anisotropy in the free-carrier transport arises from a different plasma frequencies, 2500 cm$^{-1}$ for the c-direction and 11000 cm$^{-1}$ for the ab plane, the scattering being isotropic. Interpreted as ordinary transport this would imply a mass anisotropy of $(11000/2500)^2 \approx 20$ a figure that is much larger than what is found in other measurements, for example from the anisotropy of the lower critical field [36,37] ($\approx 5$). Another problem is the source of the large residual resistance seen in the most highly-doped material. It is clear from the ab-plane transport in the same crystals, measured with microwaves [38], that the impurity scattering is extremely small in these samples and cannot account for the large intercept.

While in the normal state the one and two-component models represent different interpretations of the data, at some level they may be equivalent since the mid-infrared band could be associated as the incoherent sideband due to the interactions between the carriers. The models take a contrasting view of the absorption in the superconducting state. In the application of the two-component model to the ab plane conductivity, where clean-limit conditions apply, the residual absorption in the superconducting state is due to the mid-infrared absorption which is revealed when *all* of the Drude carriers condense. The two-component picture is a phenomenological model and the nature of the midinfrared band is not specified. In the one-component picture the residual absorption in the superconducting band is the shake-off absorption or Holstein sideband with an onset at $2\Delta + \hbar\Omega$ where $\Delta$ is the superconducting gap, and $\hbar\Omega$ the energy of the excitation that is responsible for the scattering.

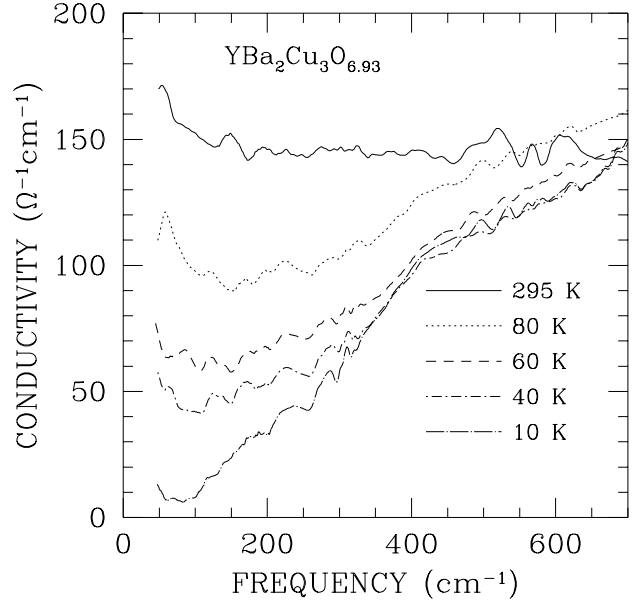

FIG. 8. The optical conductivity along the c-axis of optimally-doped $YBa_2Cu_3O_{6.93}$ (with the phonons removed) at 295, 80, 60, 40 and 10 K. The low-frequency conductivity is steadily decreasing below $T_c$ as the spectral weight is transferred to the condensate at the origin. At 10 K, the conductivity is still non-zero down to the lowest measured frequency, at $\approx 50$ cm$^{-1}$. However, the low-temperature conductivity does not approach the origin monotonically, suggesting either quadratic behavior at low frequency, or the possibility of a finite gap.

The actual conductivity in the superconducting state, shown in Fig. 8, in the fully-doped material rises approximately linearly from a very low value at low frequency to cross the normal-state value at $\approx 700$ cm$^{-1}$. The behavior of the low-temperature conductivity along the c-axis is dramatically different from the ab plane in samples from the same source, [38,39] where there is a large amount of residual absorption down to the lowest measured frequency.

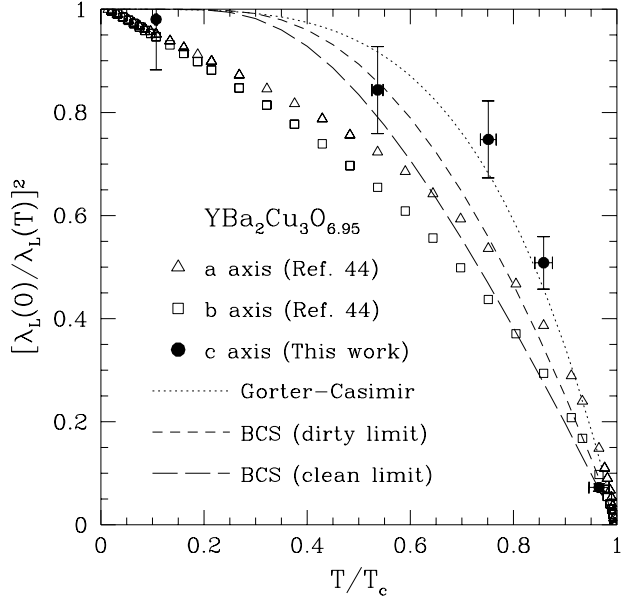

FIG. 9. The normalized London penetration depth along the c-axis of optimally-doped YBa$_2$Cu$_3$O$_{6.95}$ obtained from $\epsilon_1$, as shown by the solid circles. The open triangles and squares are the penetration depths along the a and b axes, respectively (Ref. 44), measured with microwaves. The dotted line shows the function $1 - (T/T_c)^4$, from the Gorter-Casimir two-fluid model, while the dashed lines show the BCS predictions for a dirty and clean limit superconductor.

The superconducting-state conductivity of the highly-doped samples resembles that calculated for models with nodes in the gap. [40–42] There is continuous absorption down to the lowest frequency and the temperature dependence is in agreement with models with low-lying states as well: for example, the minimum in the conductivity moves to higher frequency with temperature instead of closing as one sees in BCS superconductors with an isotropic gap. Before concluding that the c-axis spectra support superconductivity models with nodes in the gap function, we should note that our results are also consistent with a low-lying gap of the order of 80 cm$^{-1}$ or less.

The London penetration depth and the density of the superfluid condensate can be calculated from the frequency dependence of the real part of $\epsilon_1$ since for a delta function at the origin:

$$\epsilon_{1s} = 1 - \frac{\omega_{ps}^2}{\omega^2}$$

where $\epsilon_{1s}$ is the contribution to the dielectric function of the superconducting condensate, and $\omega_{ps}$ the plasma frequency of the condensate. The plasma frequency of the condensate can be then obtained from the slope of the plot of the dielectric function (with the phonons removed) as function of $1/\omega^2$. In such a plot the contribution of the carriers that have not condensed gives rise to deviations from a straight line. However, such deviations are usually small. A more serious problem is the presence of a narrow component of normal carriers with a Drude width below the lowest far-infrared frequency. The contribution to $\epsilon_1$ of such a component cannot be distinguished from that of the superconducting carriers. In the ab-plane response there is a narrow normal component, [38] and it gives rise to the linear variation of the penetration depth seen at low temperature. [43] As we have shown above, in the c-direction, the coherent Drude component is quite broad in the normal state, and we see no evidence of a narrowing of the quasi particle peak just below $T_c$, such as the one that is seen in the infrared spectra in the ab plane reflectance. [44] Nevertheless we cannot rule out the existence of such a peak and its presence can only be detected with lower frequency spectroscopy. Also, it should be noted the absolute value of the London penetration depth reported by infrared spectroscopy includes, in addition to the true condensate response, that of any other modes with frequency below the far-infrared range.

Fig. 9 shows the square of the normalized London penetration depth for the fully-doped sample obtained from an analysis of $\epsilon_1$ (solid circles). As a check we have also evaluated the loss of spectral weight in the far-infrared, and find the same values to within the error bars shown. The c-axis penetration depth reaches its low-temperature value faster than it does along either the a or b axes (the open triangles and squares, respectively). [45]

### 2. Underdoped YBCO

The removal of oxygen to produce the under-doped materials results in startlingly different behavior, as the normal-state conductivity may no longer be described as metallic with a large damping. Instead a pseudogap develops in the otherwise frequency and temperature-independent background. We have not followed the changeover from a slightly negative metallic slope with frequency seen in the $x = 0.95$ material to the constant and slightly positive slope in the underdoped cases. However, Fig. 3 suggests that this crossover occurs near $x \approx 0.92$. The exact determination of frequency dependence of the background conductivity is difficult because of the interference by the phonons and the pseudogap.

The most striking feature of the c-axis conductivity in the underdoped materials is the development of the pseudogap with an onset at $\approx 200$ cm$^{-1}$ shown in Fig. 5(a). Figure 5(b) shows in greater detail the optical conductivity of of the $x = 0.70$ material with the phonons removed. A version of these curves was published in Ref. 9, but in that work the broad phonon at 400 cm$^{-1}$ was not removed from the spectra. The shape of the pseudogap seen so clearly in Fig. 5(b) suggests that a reliable estimate for the value of the pseudogap in other dopings may be obtained by fitting the electronic continuum to a broadened step function of the form

$$\sigma_c(\omega) = \sigma_0 + \frac{\sigma_\infty - \sigma_0}{1 + \exp[(2\Delta_p - \omega)/\Gamma_p]}, \quad (4)$$

where $\sigma_0$ is the residual conductivity at low frequency, $\sigma_\infty$ is the high-frequency conductivity (in the absence of other excitations) and $2\Delta_p$ and $\Gamma_p$ are the energy and damping of the pseudogap, respectively. Figure 10 shows the low-temperature conductivity for four dopings $x \leq 0.8$ where the pseudogap can be resolved. The values for the pseudogap, while decreasing slightly with doping, are essentially the same for all of the doping levels (within error), $2\Delta_p \approx 280 \pm 20$ cm$^{-1}$. The position of the pseudogap also displays little temperature dependence. Note that because the gap is rather broad (generally, $\Gamma_p \approx 30$ cm$^{-1}$), the onset of absorption is considerably lower, in most materials at $\approx 200$ cm$^{-1}$. Except for the onset region, the conductivity below $2\Delta_p$ is frequency independent. In the cases where the pseudogap is well defined in the optical conductivity ($x \lesssim 0.8$), the pseudogap acts simply to remove spectral weight in what is otherwise a frequency-independent band.

There is a simple interpretation of a spectrum where spectral weight is simply removed from low frequencies such as the one shown in Fig. 5. In a system with two components, one of which has a flat spectrum and the other the same flat spectrum but completely gapped, one expects partially-filled gaps if the ratio of the concentration of the two components changes with temperature. At high temperature, the material would largely consist of the ungapped portion, and at low temperature the gapped portion. The gapped portion would have to have a high-frequency excitation that recovers the spectral weight from the gap. This simple picture is consistent with our conductivity spectra.

The gapped portion of the spectrum has some unusual properties. First the size of the gap appears to be temperature independent. Pairing gaps in mean-field theory usually close when the temperature reaches 30% of the gap value and strong-coupling gaps at even lower temperatures. The pseudogap in the YBa$_2$Cu$_3$O$_{6.60}$ still has its full value of 280 cm$^{-1}$ at 300 K. A second point is the apparent isotropy of the gap. Below the onset at $\approx 200$ cm$^{-1}$ the conductivity is flat at all temperatures, a signature of an isotropic gap. Third, the overall shape of the gap is reminiscent of BCS dirty-limit gaps where the conductivity rises smoothly from the threshold and does not overshoot the normal-state value. In contrast, the spin-wave gap in the heavy-fermion compound URu$_2$Si$_2$ has a peak in conductivity just above the gap value, and the excess area under that peak contains much of the missing spectral weight in the gap region.

Another feature that distinguishes the under-doped materials from the optimally-doped ones is the evolution of the spectral weight with temperature. Figure 11(a) shows the integrated spectral weight in the $0-700$ cm$^{-1}$ region, for $x = 0.95$, as a function of temperature. As the temperature decreases, to just above $T_c$, the spectral weight at low frequency increases, this is consistent with the slight narrowing of the Drude peak as the temperature is lowered. However, below $T_c$ the spectral weight decreases rapidly as the conductivity in the gap region collapses to the delta function at the origin. The same integral is shown in Fig. 11(b) for the oxygen-reduced $x = 0.70$ material, as a function of temperature. The spectral weight in this case begins to decrease in a linear fashion from room temperature down to the lowest measured temperature, but shows no anomaly at $T_c$. In the normal state, the decrease in the spectral weight at low frequency is associated with the formation of the pseudogap, and from conductivity sum rules, the missing spectral weight must be redistributed to another frequency region. We can rule out low-lying collective modes since they would raise the low-frequency reflectance and also as Fig. 2 illustrates, at low frequency the conductivity extrapolates directly to agree with the dc conductivity, also suggesting an absence of low-lying modes. Thus it appears that the spectral weight is distributed over a range of frequencies beyond the region that we have studied here. It is difficult to locate this high-frequency absorption if it is very broad. This behavior is to be contrasted with the formation of a superconducting gap where the spectral weight from the gap region is transferred to the delta function at the origin. Of course, below the superconducting transition the spectral weight must go to the delta function, and this point will be discussed in more detail below.

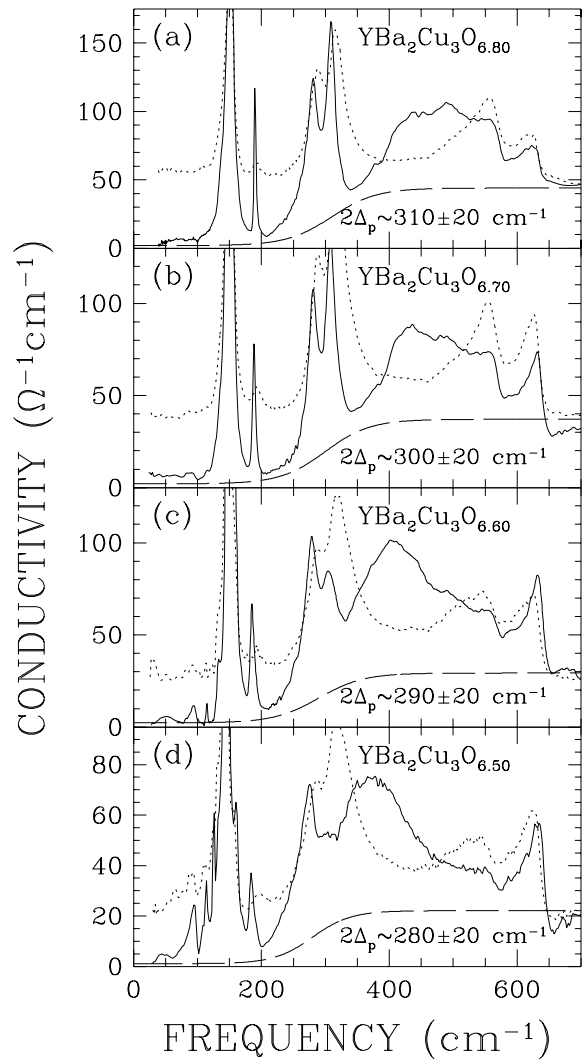

FIG. 10. The c-axis conductivity of YBa$_2$Cu$_3$O$_{6+x}$ for (a) $x = 0.80$, (b) 0.70, (c) 0.60, and (d) 0.50 at 295 K (dotted line) and 10 K (solid line), values for which the pseudogap can be resolved. The dashed line represents the best estimate of the conductivity with the phonons removed. The shape of this continuum has been modeled using a broadened step function (see text), to obtain a value for the pseudogap of $2\Delta_p \approx 280 \pm 20$ cm$^{-1}$.

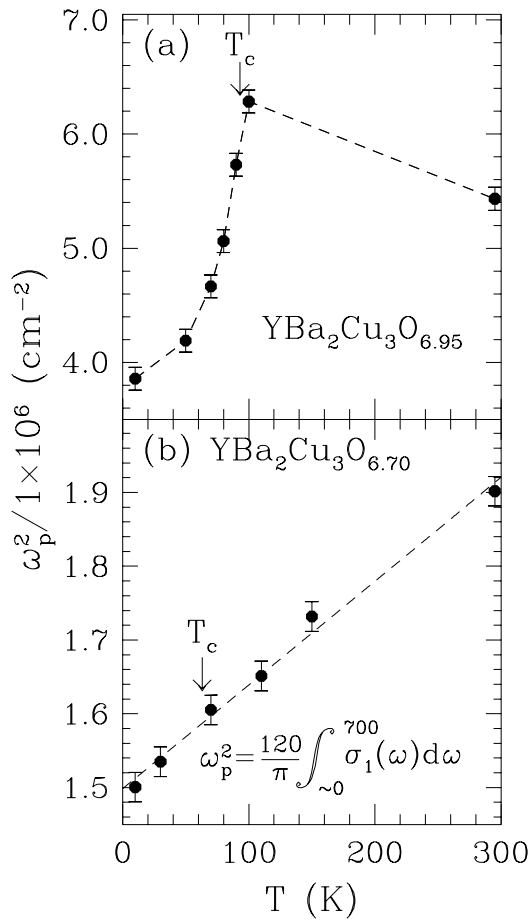

FIG. 11. The spectral weight of the conductivity, expressed as a plasma frequency, $[\omega_p^2 = 120/\pi \int_0^{700} \sigma_1(\omega)d\omega]$ of (a) $YBa_2Cu_3O_{6.95}$, and (b) $YBa_2Cu_3O_{6.70}$. Note in the upper panel, the increase in the spectral weight as the temperature decreases until $T_c$, for T<$T_c$, there is an abrupt drop as the spectral weight shifts to the superconducting condensate. In contrast, the spectral weight in the lower panel for the oxygen-reduced sample shows a steadily decreasing spectral weight, with no anomaly at $T_c$.

A careful examination suggests that the spectral weight for the superconducting condensate originates from a spectral region extending up to 1000 cm$^{-1}$. While this can be seen to be the case in both the $x = 0.60$ and $x = 0.70$ samples of $YBa_2Cu_3O_{6+x}$, it is particularly clear in the $YBa_2Cu_4O_8$ materials. [16] The following model would be consistent with these observations. Above $T_c$ the spectral weight from the pseudogap region is transferred to very high frequencies beyond our region of observation. At the superconducting transition the growth of the pseudogap ceases and the spectral weight for the condensate comes from a broad region extending up to 1000 cm$^{-1}$. That there is a change in the rate of formation of the pseudogap at the superconducting transition is also shown by the inflection point seen in the Knight shift curve at $T_c$ as well as our low frequency conductivity. It is remarkable that the slope of the spectral weight curve is smooth through $T_c$, suggesting that there is a close relationship between the superconducting gap and the pseudogap.

The idea that underdoped high-temperature superconductors have a gap in the normal state is not a new one. Nuclear magnetic resonance experiments have been discussed in terms of a spin gap which is invoked to account for the reduction of the Knight shift well above the superconducting transition in underdoped material. Since the Knight shift, in ordinary metals, is proportional to the density of states at the Fermi surface, a reduction in the Knight shift has been interpreted as evidence for the development of a gap. The inset in Fig. 5(b) compares the conductivity at $\approx 50$ cm$^{-1}$ to the Knight shift in an oxygen-reduced $x = 0.63$ sample [46] (both of which are normalized): the two are in excellent agreement. The dc transport in the ab plane in the underdoped samples shows a change of curvature at temperatures well above the $T_c$, and has been interpreted in terms of a spin gap. [47,48] Further evidence for the a gap in the normal state comes from the measurement of specific heat, [49] where a gap has been found to open above the superconducting transition temperature. The magnitude of the gap depends on the doping level, the gap gets larger as the doping is reduced from the fully-doped value, where the gap cannot be distinguished from the superconducting gap. In the $x = 0.73$ material the gap is $115 \pm 15$ cm$^{-1}$. This number agrees with our value of gap at 10 K of $280 \pm 20$ cm$^{-1}$ if our gap is interpreted as a pairing gap where the threshold for the optical absorption is $2\Delta_p$, but the specific heat follows $\exp(\Delta_p/k_BT)$ law at low temperatures.

## V. CONCLUSIONS

In the highly-doped materials, the normal-state conductivity along the c-axis shows evidence of coherent transport. The normal-state conductivity is described quite well by a simple Drude model; the one and two-component models, which are applicable in the ab plane, yield either unphysical results (in the case of the former), or results which disagree with the larger body of work (as in the case of the latter). The presence of coherent transport, and by implication a Fermi surface, means that it should be possible to observe the effects of a gap forming at the Fermi surface below $T_c$. The observation of finite conductivity down to the lowest measured frequency, and the large residual conductivity in the ab plane, suggests that the gap has nodes at the Fermi surface. However, our results are also consistent with a low-lying gap of the order $\approx 80$ cm$^{-1}$ or lower.

In the oxygen-reduced materials, several remarkable new properties are observed. The first is the crossover from coherent to incoherent transport at dopings below $\lesssim 0.90$, and the destruction of the Fermi surface along $k_c$. The second interesting feature is that while the response of the low-frequency conductivity is now non-metallic, it is very different from simple hopping conductivity. For dopings below $\lesssim 0.8$, a pseudogap develops in the normal-state conductivity and persists into the superconducting state, with no sign of an anomaly near $T_c$. The energy of the pseudogap, $2\Delta_p \approx 280$ cm$^{-1}$, shows little temperature dependence, and only a weak doping dependence (if any). The removal of spectral weight via the pseudogap suggests two components for the conductivity, one of which has a flat spectrum, and the other the same flat spectrum, but completely gapped. The agreement of the far-infrared conductivity with the Knight shift in the oxygen-reduced materials further supports the observation of a pseudogap in the normal state.

The behavior of the optical conductivity as a function of doping suggests that in the oxygen-reduced systems, there are two separate and distinct phase transitions that occur: a pseudogap at $T \gtrsim 300$ K, and a superconducting transition at low temperature.

## ACKNOWLEDGMENTS

We would like to thank P.W. Anderson, D.N. Basov, A.J. Berlinsky, J.P. Carbotte, V. Emery, C. Kallin, P.A. Lee, K. Levin, A.J. Millis, D. Pines, and J.S. Preston for stimulating discussions. This work was supported by the Natural Sciences and Engineering Research Council of Canada and the Canadian Institute for Advanced Research.

* Present address: Department of Physics, Simon Fraser University, Burnaby, B.C. Canada V5A 1S6, Email: homes@sfu.ca
** Corresponding author, Tel: (905) 525-9140, Ext. 24290, FAX: (905) 546-1252, Email: timusk@mcmaster.ca